\documentclass[aps,prl,amsmath,amssymb,groupedaddress,twocolumn,showpacs]{revtex4}
\usepackage{amssymb}
\usepackage{graphicx}

\begin{document}

\title{Competing interactions in arrested states of colloidal clays}

\author{B.~Ruzicka$^{1}$, L.~Zulian$^{2}$, E. Zaccarelli$^{1}$, R. Angelini$^{1}$,
 M. Sztucki$^{3}$, A. Moussa\"id$^{3}$, and G.~Ruocco$^{1}$}
\affiliation{ $^{1}$ SOFT INFM-CNR and Dipartimento di Fisica,
Sapienza
Universit$\grave{a}$ di Roma, I-00185, Italy.\\
$^{2}$ ISMAC-CNR via Bassini 15, 20133 Milan, Italy.\\
$^{3}$ European Synchrotron Radiation Facility. B.P. 220 F-38043
Grenoble, Cedex France.}

\date{\today}
\begin{abstract}

Using experiments, theory and simulations, we show that the
arrested state observed in a colloidal clay at intermediate
concentrations is stabilized by the screened Coulomb repulsion
(Wigner glass). Dilution experiments allow us to distinguish this
high-concentration disconnected state, which melts upon addition
of water, from a low-concentration gel state, which does not melt.
Theoretical modelling and simulations reproduce the measured Small
Angle X-Ray Scattering  static structure factors and confirm the
long-range electrostatic nature of the arrested structure. These
findings are attributed to the different timescales controlling
the competing attractive and repulsive interactions.

\end{abstract}
 \maketitle

Dynamical arrest in soft colloidal systems has recently become the
subject of an intense research activity. The fine tuning of
control parameters opens the possibility to tailor the macroscopic
properties of the resulting non-ergodic states. Several mechanisms
of dynamical arrest have been identified. Building on the
knowledge on the hard spheres glass~\cite{Pusey},
 it has become recently clear
that when both attractive and repulsive terms are present in the
interaction potential, a re-entrant liquid-glass line, surrounded
by two distinct glasses, has been predicted and experimentally
observed in short-ranged attractive colloids at high
concentrations~\cite{SciortinoNatMat}. A rich phenomenology also
takes place at low concentrations: here gelation occurs, which may
result from different routes~\cite{ZaccarelliRev}. Interesting
scenarios arise when, in addition to a short-ranged attraction,
particles have a residual electrostatic charge which builds up a
long-range repulsion in the effective potential. In this case,
particles can form equilibrium clusters~\cite{Groenewald}, which
provide the building blocks of arrest~\cite{Sciortino}. Recent
works have shown that both Wigner glasses~\cite{Cha1982}, intended
as arrested states formed by disconnected particles or clusters
and stabilized by the electrostatic repulsion, and equilibrium
gels, which occur at larger packing fractions when the clusters
branch into a percolating network, can form under these
conditions~\cite{ZaccaSoftMatter, Royall}.

To investigate the formation of multiple arrested states,
colloidal clays~\cite{Shalkevich,Lekkerkerker} have emerged as
suitable candidates. The anisotropy of the particles, combined
with the presence of attractive and repulsive terms in the
interactions, makes the phase diagram of such colloidal systems
very complex. Among these, Laponite suspensions have been widely
studied not only for their appealing industrial
applications~\cite{Laporte} but also for their interesting
experimental/theoretical properties~\cite{Mourchid, Bonn,
Mongondry1,
Ruzicka,Schosseler,JabbariPRE2008,RuzickaPRE2008,Trizac, Mossa}.
In particular, Laponite displays a non-trivial aging
dynamics~\cite{Ruzicka} and (at least) two final arrested states,
which are obtained by a simple increase of Laponite volume
fraction from low ($C_w < 2.0 \%$) to moderate ($C_w \geq 2.0 \%$)
values, at fixed salt concentration $C_s=10^{-4}M$~\cite{Ruzicka,JabbariPRE2008}.
More recently, the static properties of these two states have been
investigated in detail~\cite{RuzickaPRE2008}, showing that they
are also characterized by both the very different aging behaviors
and the shape of the static structure factor $S(Q)$. While the low
$C_w$ state shows typical inhomogeneities, the high $C_w$ state is
homogenous, allowing for the identification of  these two arrested
states respectively as gel and glass.

\begin{figure*}[t]
\centering
\includegraphics[width=.8\textwidth]{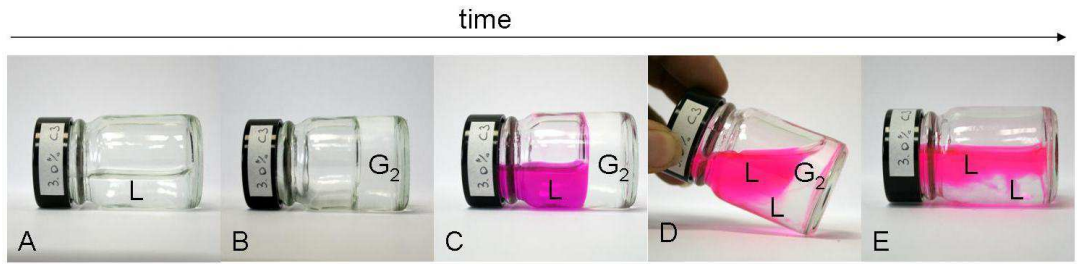}
\caption{Evolution of a high concentration sample at $C_w=3.0 \%$
with time. The initially liquid ($\bf L$) sample (A) ages with
time reaching, after 50 h, a final arrested glassy state (${\bf
G_2}$)(B). At this time a colored solution (in a liquid ($\bf L$)
state) is added to the arrested sample (C) that starts to fluidize
at the interface (D)- 70 h. Finally the whole sample is liquid
again (E) - 120 h.} \label{Fig1}
\end{figure*}

In this Letter, through the combination of experiments, numerical
and theoretical approaches, we demonstrate that the non-ergodic
state observed in Laponite samples in salt free water at
sufficiently large concentrations ($C_w \geq 2.0 \%$) is a Wigner
glass. By performing a simple but impressive dilution experiment,
we are able to distinguish whether attractive or repulsive
interactions are dominant in the formation (and stability) of the
arrested structure for both low and high concentration arrested
samples. Moreover, we compare the  $S(Q)$ determined by Small
Angle X-rays Scattering (SAXS) with theoretical and numerical ones
in the high concentration window, providing an estimate for the
effective interactions between Laponite platelets. These results
provide clearcut evidence that the resulting large concentration
arrested state (which is still found at very low packing fraction)
can be attributed to the residual electrostatic repulsion (Wigner
glass)~\cite{Cha1982}. Thereby Laponite offers the unique example
of a system which, not only displays two distinct non-ergodic
states differing only in colloidal concentration, but also shows a
counter-intuitive scenario which involves gel-like structures at
lower concentrations and a (truly) disconnected glass at larger
ones. Hence, the increase of concentration, which normally favours
percolation and branching, here has the opposite effect. We
attribute this novel behavior to the interplay of different
time-scales in the involved attractive and repulsive interactions,
a feature that may be found in other complex colloidal systems.

To investigate the glassy state formed by Laponite at high clay
concentration ($C_w\geq 2.0
\%$)~\cite{Ruzicka,JabbariPRE2008,RuzickaPRE2008}, we use a
combination of different approaches. Firstly, we perform a
dilution experiment (adding water to the arrested state) that
provides a direct criterion to establish the dominant interactions
controlling the formation and stability of the arrested
state~\cite{ZaccarelliRev}. If this were due to attractive
interactions, the presence of additional water should not affect
the final state, because Laponite bonds (of strong electrostatic
nature) can not be broken by the mediation of water. On the other
hand, if repulsive interactions are determining the formation of
the non-ergodic state, the increase of free volume allows a
rearrangement of the Laponite platelets on average to a larger
distance. If such distance is larger than the characteristic
repulsive length (namely the electrostatic Debye length) the glass
will be destabilized up to melting to a liquid state.

To obtain reliable and reproducible results~\cite{CumminsJNCS2007}
Laponite RD dispersions are prepared using the same protocol
described in~\cite{RuzickaPRE2008}. The starting waiting time
($t_w=0$) is the time at which the suspension is filtered. For the
dilution experiment a sample at high clay concentration, $C_w=3.0
\%$, is filtered and sealed in small glass bottles. The
photographic sequence of this experiment is shown in
Fig.\ref{Fig1}. The initially liquid (${\bf L}$)((A) panel) sample
ages with time $\approx 50$ h~\cite{Ruzicka} until reaching a
final arrested state, as it can be seen in panel (B) where the
sample shows its solid like nature (${\bf G_2}$). At this point a
solution of colored water (of the same volume as that of the
Laponite solution) is added to the arrested (colorless) sample.
The colorant used, Rhodamine B at a concentration $10^{-3} M$,
does not play any role in the process, as shown below, and is used
to better distinguish (C panel) the arrested sample (${\bf G_2}$)
from the liquid solution (${\bf L}$). After the addition of water,
the arrested sample starts to fluidize at the interface with the
liquid solution (D panel - $t_w=70$ h). This process evolves with
time until, after 120 h, the whole sample is back into a liquid
state (E panel). This result strongly speaks in favor of the
repulsive nature of the $C_w= 3.0 \%$ arrested state. The
experiment is repeated in absence of Rhodamine and shows that the
addition of the dye does not affect the results. We have monitored that the pH of the solution remains constant throughout the dilution experiment.

We now turn to elucidate the effective interactions between
Laponite platelets, by comparing the evolution of the experimental
static structure factor $S^M(Q)$, obtained from SAXS measurements
performed at the High Brilliance beam line (ID2) \cite{Narayan} at
the European Synchrotron Radiation Facility (ESRF) in Grenoble,
France, as described in~\cite{RuzickaPRE2008}, with that predicted
theoretically $S^{th}(Q)$, assuming several effective interaction
potentials \cite{Likos}.

\begin{figure}[t]
\centering
\includegraphics[width=7cm,angle=0,clip]{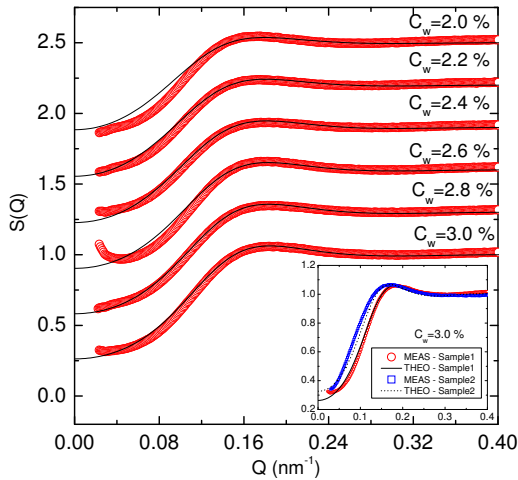}
\caption{Comparison between measured $S^M(Q)$ (symbols) and
theoretical $S^{th}(Q)$ (lines) for several high concentration
Laponite samples at $t_w\simeq 50$ h. For clarity, curves have
been shifted along the vertical axis progressively by 0.3. Inset:
Comparison between $S^M(Q)$ (symbols) and $S^{th}(Q)$ (lines) for
two different samples with $C_w=3.0 \%$.} \label{Fig2}
\end{figure}

In Fig. \ref{Fig2} the comparison between $S^M(Q)$ (symbols) and
optimal
 $S^{th}(Q)$ (lines), for fixed $t_w\simeq 50$ h in the concentration window
$2.0\% \leq C_w \leq 3.0\%$,  is shown. As it is evident from Fig.
\ref{Fig2}, as concentration is decreased from $C_w= 3.0 \%$ to
$C_w=2.0 \%$ the low $Q$ signal is slightly increased and the main
peak is slightly shifted towards lower $Q$ values. The position of
the main peak is around $Q\sim 0.17$ nm$^{-1}$, corresponding to a
length $\approx 37$ nm, a value clearly larger than the platelet
diameter $d=25$ nm, pointing to a non-connected final structure.
Moreover, no evidence of a peak at contact distance is found.

The theoretical $S^{th}(Q)$ is calculated by solving numerically
the Ornstein-Zernike equation with an appropriate closure
relation~\cite{hansen06} or by direct simulation of the effective
potential, without a significant change in the description of the
experimental data. For simplicity, we report Percus-Yevick results
in the following. In this simplified treatment, we consider only
the centers of mass of the scattering objects, supposedly single
platelets, and treat them as (spherical) points, following
previous work on different systems~\cite{ZaccaSoftMatter}. The
main fit parameters involved in the description of the
experimental data are the number density $\rho$ of the scattering
objects and the number of counter-ions $N_{ci}$ that dissociate
(on average) from each platelet when dissolved in water. Our study
indicates that $\rho$ cannot be the same as the number density of
Laponite discs obtained from the nominal weight concentration
$\rho_L=2.4\cdot 10^{22}/m^3$. Indeed, no simple interaction
potential is able to reproduce the position of the main peak using
$\rho_L$. Hence $\rho$ must be smaller, due to the fact that
platelets may be found within a distribution of
clusters~\cite{LapoAFM}. We find that the data can be fully
described, in terms of peak position, peak height and
compressibility, by a Yukawa potential, which accounts in an
average way~\cite{ZaccaSoftMatter} for the screened electrostatic
repulsion between Laponite platelets or clusters, in agreement
with more sophisticated theoretical approaches~\cite{Trizac}. We
use a potential of the form $V_{eff}=A\xi e^{-r/\xi}/r$, where
$\xi$ is the Debye screening length and $A$ provides a measure of
the strength of the repulsion.  The number density is fixed to
$\rho \simeq 0.38\rho_{L}$, a relationship which is found to hold
in the whole investigated concentration window. This suggests that
Laponite may be dispersed in a distribution of very small
aggregates, possibly due to particles that are not completely
delaminated, compatibly with an AFM study performed under very
dilute conditions~\cite{LapoAFM}. The number of counterions in
solution is optimized to $N_{ci}=60$, a value well below the bare
charge of a single platelet ($700 \ \ e$), in good agreement both
with previous simulations~\cite{Trizac} and with conductivity
measurements~\cite{JabbariPRE2008}. With this choice of
parameters, the change in nominal concentration and its associated
screening length $\xi$, varying between 8 and 10 nm, are able to
reproduce $S^{M}(Q)$ in the full investigated window, as seen in
Fig.~\ref{Fig2}. The repulsion strength $A\xi/k_BT$ is found to
 increase in the studied range of concentration by a factor of $\approx 2$,
compatibly with the behavior of other charged
systems~\cite{CardinauxEPL}.

While the Yukawa  potential correctly reproduces the experimental
data, other possible candidates do not. The simple hard sphere
model does not capture at all the softness of the interactions,
failing to reproduce the low-$Q$ regime also after one adjusts
$\rho$. On the other hand, the addition of a short-range
attraction to the electrostatic repulsion is (i) not sufficient to
explain the low $Q$-position of the main peak; (ii) tends to
induce an enhanced structuring at nearest-neighbour length.
Therefore, we conclude that no relevant attractive features are
present in $S(Q)$ in this concentration window. Hence, the
combination of the dilution experiment with the theoretical
analysis allows to identify the observed arrested state as a
Wigner glass stabilized by the residual electrostatic repulsions.

One further comment concerns the robustness and reproducibility of
our findings
 for different samples. Although different measurements
can bring small changes in $S^{M}(Q)$ due to differences in
Laponite batches, to the filtration procedure or to real clay
concentrations, we can fit equally well different samples with the
same effective Yukawa potential upon variation of the fit
parameters. This is shown in the inset of Fig. \ref{Fig2} where
$S^M(Q)$ is reported for two different samples with $C_w=3.0 \%$,
from different batches, measured in two different runs, in
comparison with $S^{th}(Q)$. For the second batch, which has a
higher polydispersity, we find $N_{ci}=90$ and $\rho=0.26 \rho_L$.

\begin{figure}[t]
\centering
\includegraphics[width=7cm,angle=0,clip]{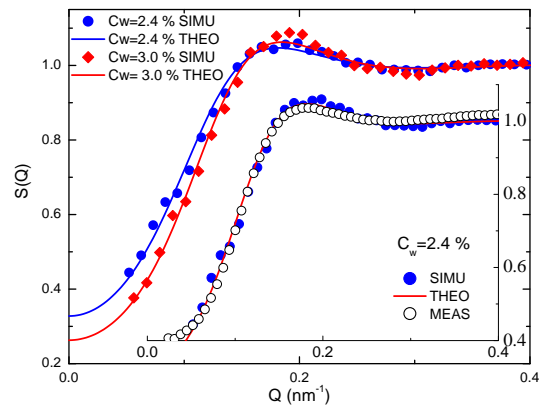}
\caption{Comparison between simulated (symbols) and theoretical
(lines) S(Q) for $C_w=2.4\%$ and $3.0\%$ Laponite samples. Inset:
Simulated (cross symbols), theoretical (line) and measured (open
circles) $S(Q)$ for $C_w=2.4\%$.} \label{Fig3}
\end{figure}

The theoretical approach we have adopted so far treats only the
correlations between the centers-of-mass of the scattering
objects. Although such an approach was proven to be a good
(although crude) approximation to describe the static correlations
between equilibrium clusters of non-spherical
shape~\cite{ZaccaSoftMatter}, it is crucial to show that this
treatment is valid also when we take into account explicitly the
actual disc shape of clay particles. To this aim, we have
performed MonteCarlo simulations of 200 discs at the same $\rho$
as that extracted from the fits. We use the interaction model
introduced in~\cite{HansenSimu} where the total negative charge is
uniformly distributed over the surface area, while rim charges are
neglected. Each Laponite platelet is schematized as a rigid disk,
composed by 19 sites disposed on a regular mesh. We simulate
several repulsion parameters and find that the numerical
$S^{n}(Q)$ is in good agreement with the experimental and
theoretical ones when the total charge is fixed to $70 \  e$ and
the screening length is 5 nm. Results are shown in Figure
\ref{Fig3} for two samples with $C_w= 2.4 \%$ and $C_w= 3.0 \%$,
in comparison with $S^{th}(Q)$. In the inset, simulated,
theoretical and measured $S(Q)$ for sample $C_w= 2.4 \%$ are
compared. Hence, the favorable comparison of simplified theory and
simulations with experimental measurements provides evidence that
our effective approach is useful to describe disc-shaped objects
in this regime.

So far we have reported evidence of the existence of a Wigner
glass state in Laponite by the combination of experiments, theory
and simulation for $2.0 \% \leq C_w \leq 3.0 \%$. At lower
concentrations structural measurements have shown a marked growth
of $S(Q)$ at low $Q$ with waiting time, which was attributed to
the presence of attractive interactions determining an arrested
state of attractive (or gel) nature~\cite{RuzickaPRE2008}. Already
for the lowest concentration shown here ($C_w=2.0 \%$), the
agreement between $S^M(Q)$ and $S^{th}(Q)$ is less good in the low
$Q$ range, as visible in Fig. \ref{Fig2}. This might be an
indication of the increasing role of attractive interactions upon
lowering concentration. Hence, we perform the dilution experiment
also for a low concentration sample expecting a different result
with respect to the high concentration one.

\begin{figure}[t]
\centering
\includegraphics[width=.5\textwidth]{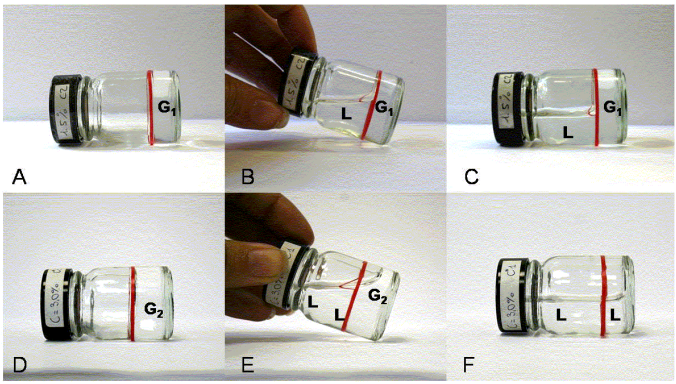}
\caption{Starting (A), intermediate (B) and final (C) states of a
low concentration sample at $C_w=1.5 \%$: the initial gel state
${\bf G_1}$ is not macroscopically affected by the addition of
water (C). Starting (D), intermediate (E) and final (F) states of
a high concentration sample at $C_w=3.0 \%$: after the addition of
water ($\bf L$) the original glassy state ${\bf G_2}$ fluidizes
(F).} \label{Fig4}
\end{figure}

In Fig.~\ref{Fig4} the dilution experiment is shown for a low
($C_w=1.5 \%$) and a high ($C_w=3.0 \%$) concentration sample. The
samples are left to age up to the final corresponding arrested
states, respectively of gel ($\bf{G_1}$) and glass ($\bf{G_2}$)
nature for low and high concentrations~\cite{RuzickaPRE2008}.
While the sample at $C_w=3.0 \%$ arrests within 50 h, the one for
$C_w=1.5 \%$ takes several weeks~\cite{Ruzicka,JabbariPRE2008}.
After arrest takes place deionized water ($\bf L$) is added and
panels (A) and (D) show the identical situation for the two cases.
However the evolution of the two samples soon becomes dramatically
different. The arrested state of the high concentration sample
($\bf{G_2}$) starts to fluidify ((E) panel) up to reach a final
liquid state ((F) panel). On the contrary, the low concentration
sample does not liquify and does not show any macroscopic changes
of its solid like state, even after waiting additional weeks, i.e.
the same timescale of the arrest process. In this case the
presence of available free volume does not determine any change in
the gel-arrested state ((B) and (C) panels), due to the fact that
water cannot break clay bonds. The different behavior between the
high and low concentration samples ((C) and (F) panels) observed
in the dilution experiment confirms the different nature of the
corresponding arrested states: a Wigner glass ($\bf{G_2}$) and  a
gel ($\bf{G_1}$), dominated respectively by repulsive and
attractive interactions at high and low concentrations.

The striking result of a low-concentration gel network and of a
high-concentration disconnected Wigner glass is at odds with
previous studies of simpler systems~\cite{Bartlett,
ZaccaSoftMatter, Royall}. We attribute this unexpected scenario to
the separate timescales controlling the interactions and the two
arrest processes \cite{Ruzicka}. While repulsion is felt almost
immediately after samples are prepared, attraction, due to its
anisotropic nature and to the presence of an effective repulsive
barrier, develops on a much longer timescale as in
reaction-limited aggregation. Long-time attraction may also affect
the repulsive Wigner glass, through the formation of subsequent
additional bonds. Indeed, rheological measurements for rejuvenated
samples show a dependence on the idle time waited after the glass
is formed~\cite{JoshiCondMat2009}. To elucidate this point we have
performed an additional dilution experiment on the high
concentration sample waiting one week after dynamical arrest takes
place. In this case the system does not melt, probably due to
intervening long-time attraction which strengthens the glass.
However the unmelted final state is different from that observed
at low concentration, while the latter remains unchanged and water
does not penetrate into it, the former seems to slowly absorb
water in the bulk. SAXS measurements show that $S(Q)$ does not
change significantly in this time window demonstrating that the
disconnected structure of the repulsive glass is preserved despite
the intervening attraction, and hence the resulting state cannot
be considered an attractive glass in the most common sense.

In conclusion, through the combination of dilution experiments,
SAXS experiments, theory and simulations, we have shown the
presence of a disconnected Wigner glass state in a screened
charged colloidal system. This arises at a larger (but still very
low) concentration with respect to the one where a gel network is
found, thanks to the different timescales controlling the
repulsive (short-time) and attractive (long-time) interactions. We
expect these findings to be relevant to other complex liquids with
competing interactions as anisotropic patchy
colloids~\cite{Glotzer} and globular proteins~\cite{Piazza}.

We thank F. Sciortino for fruitful discussions and ESRF for
provision of beamtime. EZ acknowledges ERC-226207-PATCHYCOLLOIDS
for support.

\end{document}